\documentclass{iopart}

\usepackage{iopams}
\usepackage{graphicx}
\usepackage{url}

\newcommand{\inner}[2]{\langle #1,#2 \rangle}

\newcommand{\Hz}{{\mbox{Hz}}}

\newcommand{\Mpc}{{\mbox{Mpc}}}

\newcommand{\insp}{{\mathrm{insp}}}
\newcommand{\merge}{{\mathrm{merge}}}

\newcommand{\dt}{{\Delta t}}
\newcommand{\ring}{{\mathrm{ring}}}
\newcommand{\MSol}{{M_\odot}}
\newcommand{\TSol}{{T_\odot}}
\newcommand{\be}{\begin{eqnarray}}
\newcommand{\ee}{\end{eqnarray}}
\newcommand{\nn}{\nonumber \\}

\newcommand{\ft}[1]{\tilde{#1}}

\renewcommand{\comment}[1]{}

\begin{document}

\bibliographystyle{iopart-num}

\title{Gravitational wave detection using multiscale chirplets}

\author{E J Cand\`{e}s$\null^1$, P R Charlton$\null^2$
and H Helgason$\null^3$}

\address{$\null^{1,3}$Division of Applied and Computational
Mathematics, California Institute of Technology, Pasadena CA 91125
USA}
\address{$\null^2$School of Computing and Mathematics, Charles
Sturt University, Wagga Wagga NSW 2678 Australia}

\eads{$\null^1$\mailto{emmanuel@acm.caltech.edu},
$\null^2$\mailto{pcharlton@csu.edu.au},
$\null^3$\mailto{hannes@acm.caltech.edu}}

\begin{abstract}
A generic `chirp' of the form $h(t) = A(t) \cos \phi(t)$ can be closely
approximated by a connected set of {\em multiscale chirplets} with
quadratically-evolving phase. The problem of finding the best
approximation to a given signal using chirplets can be reduced to that
of finding the path of minimum cost in a weighted, directed graph, and
can be solved in polynomial time via dynamic programming.  For a
signal embedded in noise we apply constraints on the path length to
obtain a statistic for detection of chirping signals in
coloured noise. In this paper we present some results from using this
test to detect binary black hole coalescences in simulated LIGO noise.
\end{abstract}


\section{Introduction}
\label{sec:introduction}

Despite having achieved unprecedented sensitivities, experiments for
laser interferometric detection of gravitational waves such as LIGO
\cite{ligo:web} face significant challenges, not least of which is the problem of
detecting unmodelled or poorly modelled sources of gravitational waves.
For detecting the inspiral of a binary system, the standard technique
is {\em matched filtering} using a bank of templates parametrised by
the component masses of the system.  For low-mass binaries, the time
evolution of the inspiral is well-modelled by post-Newtonian
approximations, however for high-mass binaries the models are
considerably less certain \cite{dis:comp}. Furthermore, as the binary
mass increases, the spin of the two bodies becomes a significant
factor in the evolution of the signal \cite{acst:spin}. A complete
description of a binary system including the spin of both bodies
requires $17$ parameters, making the set of templates to be searched
over infeasibly large. Even when some parameters are neglected,
estimates of the number of templates needed to detect, for example,
spinning extreme mass ratio inspirals using a space-based detector
such as LISA range from $10^{15}$--$10^{40}$ templates
\cite{ss:templates,gbcclpv:event}. Methods have been proposed to
reduce the number of templates required, such as by using detection
template families which cover the expected range of gravitational
wave signals \cite{bcv:detecting,pbcv:physical}, but these still
require $\sim 10^5$\ templates \cite{bcptv:detecting}.

Template methods for detecting binary coalescence events mostly focus on the
inspiral component or the ringdown component \cite{lg:search} and do
not attempt to match the merger component, believed to be a major
contribution to the gravitational signature for
black hole coalescences. Modelling the inspiral and ringdown is
relatively straightforward, whereas modelling the merger requires robust
techniques for solving the full Einstein equations numerically under
extreme conditions. Much progress has been made in achieving this goal
but the problem is far from solved \cite{fp:evolution,bcckm:binary}.

A number of potential gravitational wave signals are of short duration
(less than 1 second) and are collected under the heading of {\em burst
sources}. These include events such as supernovae, the final stages of
binary black hole coalescence, and other potential sources of
gravitational waves such as gamma-ray bursts. Generally, models for
these sources are either non-existent or insufficient for constructing
matched filters, and we must rely on {\em non-parametric}
methods. Various methods for detecting bursts have been proposed
\nocite{ab:tf,abcf:ep,js:time,km:wb,sc:bursts,cp:best}
\cite{ab:tf}--\cite{cp:best}, and some have been 
applied to interferometer data \cite{lsc:burstsS4}.

In this paper we apply a non-parametric detection scheme called the {\em best
path (BP) test} introduced in \cite{chc:theory} to the detection of
binary black hole coalescences in simulated LIGO noise.  The terminology
comes from the study of weighted graphs and refers to the path between
two vertices of a graph which is of maximum total weight, subject to a
constraint on it's length.  The BP test is applicable to the
detection of quasi-periodic signals of the form
\be h(t) &=& A(t) \cos \phi(t)
\label{eq:prototype}
\ee
where the amplitude $A(t)$ varies slowly with time and the unknown
phase $\phi(t)$ obeys some regularity conditions. Signals of this form
have a well-defined {\em instantaneous frequency} $f(t) =
\dot{\phi}(t)/2\pi$ (to avoid confusion, we note that there is an
unrelated method called the Fast Chirp Transform which is
applicable to the detection of signals of the form (\ref{eq:prototype})
where the phase function is known \cite{jp:fct}).


\section{Chirplet path pursuit}
\label{sec:method}

Given detector output
\be
u(t) &=& n(t) + \rho\,h(t)
\ee
where $n(t)$ is Gaussian coloured noise with $2$-sided power spectral density
$S(f)$, we seek a test statistic which will discriminate between
the two hypotheses 
\be
H_0 &:& \rho = 0 \nn
H_1 &:& \rho \ne 0\,.
\ee
The null hypothesis is that the data is pure noise, while the
alternative is that the data contains a chirp-like signal
of the form (\ref{eq:prototype}), normalised with respect to the inner
product derived from $S(f)$,
\be
\inner{u}{v} &=&
\int_{-\infty}^{\infty} \frac{\ft{u}^*(f) \ft{v}(f)}{S(f)}\, df\,.
\label{eq:inner}
\ee
The parameter $\rho$ may be interpreted as the expectation of the SNR,
\be
\mbox{SNR} &=& \frac{\inner{u}{h}}{\mbox{rms}\,\inner{n}{h}}\,.
\ee
Note that a $1$-sided PSD is more commonly used in the 
literature, equivalent to $2 S(|f|)$. We use the $2$-sided PSD here to simplify
the discretised form of (\ref{eq:inner}).

Locally, chirps with smoothly-varying phase have a very simple
structure. Over short times their frequency evolution is approximately
linear. For longer duration, local approximations can be joined
together so that the instantaneous frequency of the signal is
approximated by a piecewise linear function.  In the following we
outline the methodology for obtaining a test statistic via chirplet
path pursuit -- details may be found in \cite{chc:theory}.

\subsection{Multiscale chirplets}

Consider a signal on the interval $I = [0, T)$. The preceding discussion
suggests we should examine functions which will correlate well locally
with signals of the form (\ref{eq:prototype}).
Our detection method uses a dictionary of normalised {\em multiscale chirplets}
of the form
\be
c_{s,j,a,b}(t) &\propto& e^{i 2\pi(a t + b t^2/2)}, \qquad t \in I_{s,j} \subseteq I
\ee
that is, a collection of chirplets supported on intervals $I_{s,j}$
and parametrised by length scale $s$, location $j$, initial frequency $a$ and chirp rate $b$.
The intervals are taken to be {\em dyadic} of the form $I_{s,j} = [j 2^{-s}T, (j+1) 2^{-s}T]$.
Here $s = 0, 1, 2, \ldots$ represents a scale index and defines the length
of the dyadic interval. The dictionary has elements of various
durations, locations, initial frequencies and chirp rates. It is convenient to
think of a chirplet as a line segment $a + b t$ supported on $I_{s,j}$ in the
time-frequency plane.

Our test statistic is constructed by looking for a connected `path' of
chirplets in the time-frequency plane that gives a good overall
correlation with the signal. To achieve this we notionally discretise
the time-frequency plane and consider points $(t_i, f_k)$ as vertices
in a directed graph. The frequency
intervals may be chosen as convenient -- for example, to coincide with
bins of a discrete Fourier transform.  Fixing a time-frequency
discretisation also fixes the the discretisation of the chirp
parameter, since we think of chirplets as arcs connecting vertices of
the graph supported on dyadic intervals. Using the FFT we can
quickly calculate the local correlations $|\inner{u}{c_{s,j,a,b}}|^2$ of $u(t)$
with elements of the chirplet dictionary, which we use as the weights of
the arcs connecting each vertex in the graph.
Given a connected, non-overlapping chirplet path $P = \{c_1, c_2, \ldots, c_p\}$ supported on a partition
${\cal P} = \{ I_1, I_2, \ldots, I_p \}$ of $I$ the total weight of the path is $\sum_{p} |\inner{u}{c_p}|^2$.
A description of our discretisation scheme may be found in the Appendix.

Simply maximising $\sum_{p} |\inner{u}{c_p}|^2$ over all chirplet
paths will naively overfit the data. In the limit of small chirplets, such
a statistic would simply fit $u(t)$ rather than a hidden signal. Instead
we use a {\em multivariate} statistic obtained as the solution of the optimisation
problem
\be
T^*_\ell &=& \max_{P} \ \sum_{p} 
|\inner{u}{c_p}|^2 \quad\mbox{subject to}\quad |P| \le \ell.
\label{eq:TstarEll}
\ee
Here $\ell$ is a constraint on the path length ie. the number of
chirplets in the path.  To be adaptive, we calculate $T^*_\ell$ for
several different path lengths, $\ell \in L = \{ \ell_1, \ell_2,
\ldots \}$. While there are a vast number of possible paths, using a
variant of Dijkstra's algorithm, calculating $T^*_\ell$ reduces to a
constrained dynamic programming problem which can be solved in
$\Or(|L|\times \#\mbox{arcs})$ \cite{hj:shortest}. The number of arcs
depends on such things as choice of discrete frequencies and chirp rates, but is
typically not more than $N^2 \log_2 N$.

\begin{figure}[ht]
\begin{center}
\includegraphics[width=0.5\textwidth]{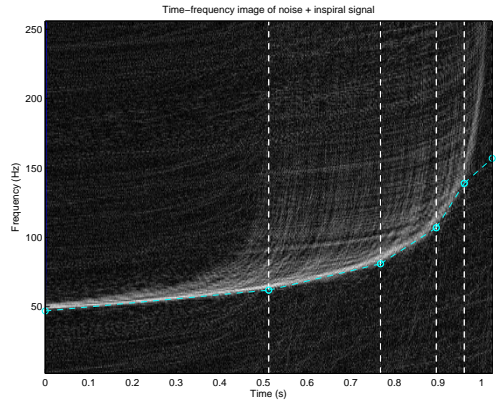}
\end{center}
\caption{\label{fig:bp_demo}
Best path found for a binary inspiral signal with total mass $16\ \MSol$
in white noise, indicated by the dashed curve. Vertical lines delimit
the support of individual chirplets in the path.
Notice that the BP test uses long chirplets
when the frequency is changing slowly, and short chirplets when it
is changing rapidly.
}
\end{figure}

Since $T^*_\ell$ is a multivariate statistic we use a {\em multiple
comparison} rule for rejecting the null hypothesis \cite{bh:false}.
Given data $u(t)$ we test $H_0$ at false alarm probability $\alpha$ using the following procedure:
\begin{enumerate}
\renewcommand{\labelenumi}{\arabic{enumi}.}
\item For each $\ell \in L$, calculate $T^*_\ell$ and find the corresponding
$p$-value under $H_0$, $p_\ell$.
\item Compare the minimum $p$-value $p^* = \min_\ell p_\ell$ with
the distribution of minimum $p$-values under $H_0$.
\item If $p^*$ is small enough to lie in the $\alpha$-quantile of
the distribution, reject $H_0$ -- we conclude a signal is present.
\end{enumerate}
In this procedure we are choosing the ordinate of the multivariate
test statistic that gives the greatest evidence against the null
hypothesis. We then compare this $p$-value with what one would expect
under the null hypothesis. Although there do not exist analytic
expressions for the distributions of $T^*_\ell$ and the
minimum $p$-value, we can estimate them using Monte Carlo
simulations. We call $T^*_\ell$ the {\em best path (BP) statistic}.
As an example, Figure \ref{fig:bp_demo} shows the path obtained for
an inspiral signal in white noise.


\section{Simulations}
\label{sec:simulations}

\subsection{Noise model}

To estimate the statistical power of the BP test we have studied the
detection of certain gravitational wave signals in simulated LIGO noise.
Discretely sampled Gaussian noise is produced via the
following method. We generate two sequences of white noise $a_k,\ b_k$,
then construct a discrete Fourier representation of an
instance of coloured noise $\ft{n}_k$ using the PSD as follows:
\be
\ft{n}_0 &= \left[\frac{N S_k}{\dt}\right]^\frac{1}{2}\, a_0 &\nn
\ft{n}_k &= \left[\frac{N S_k}{\dt}\right]^\frac{1}{2}\,\frac{a_k + i
b_k}{2} \qquad &k = 1, \ldots, N/2-1 \nn \ft{n}_{N/2} &= \left[\frac{N
S_{N/2}}{\dt}\right]^\frac{1}{2}\, a_{N/2} &\nn \ft{n}_k &=
\ft{n}^*_{N-k} \hfill &k = N/2 + 1, \ldots, N-1.
\ee
By construction, the inverse DFT $n_k$ is real Gaussian noise with PSD
$S_k$. The PSD used is the polynomial fit given in \cite[Table
5]{glpps:gwa}. This fit is only valid for frequencies above the LIGO-I
seismic wall frequency $f_s = 40\ \Hz$.  Seismic noise renders region
below $f_s$ inaccessible to gravitational wave searches.  For the
purposes of simulation, we mimic high-pass filtered data by rolling
off $S(f)$ below $20\ \Hz$. When calculating the BP statistics we only
search over paths with instantaneous frequencies above $f_s$. At this
time we have not included non-Gaussian features such as instrumental
bursts in our noise model.

\subsection{Signal model}
\label{sec:signal}

Since the object of the exercise is to detect `real' gravitational
waves, we will use as our test signals a collection of physically
realistic waveforms for binary black hole coalescence. We use a
modification of the method in \cite{ab:tf} to model a complete
coalescence waveform.  The signal consists of an inspiral component, a
merger component, and a ringdown component. While the inspiral and
ringdown models are reasonable, the simulated merger should not be
taken to be physically realistic. Instead, it is meant to approximate
the overall time and frequency characteristics of a real merger.

The test signals are parametrised by the total mass $M = m_1 + m_2$ of
the two bodies and the symmetric mass ratio $\eta = m_1m_2/M^2$.  The
full waveform is obtained by combining the components in such a way
that the instantaneous frequency and amplitude are continuous up to
first derivatives:
\be
h(t) &=& \left\{
\begin{array}{ll}
A^\insp(t)\cos \phi^\insp(t) & t \leq 0 \\ 
A^\merge(t) \cos \phi^\merge(t) & 0 < t \leq t_m \\
A^\ring(t) \cos \phi^\ring(t) & t_m < t\,.
\end{array}
\right.
\ee
Here we have arranged for the inspiral component to end at $t = 0$ and the
merger component to end at $t = t_m$.
Following \cite{fh:waveforms} we take
the merger duration to be $t_m = 50 M/\MSol \times \TSol$.

For the inspiral component of the signal we use the non-spinning 2PN
approximation for the phase in the form given by \cite[eqn. 15.24]{lsc:lsd}.
For amplitude we use the leading order (ie. Newtonian) expression given in
\cite[eqn. 15.27--28]{lsc:lsd}. For simplicity we average over
orientation $(\iota, \beta)$ and sky position to obtain
\be
A^\insp(t) &=& \frac{8}{5} \frac{\TSol c}{D}\frac{\eta M}{\MSol}
\left[\frac{\pi \TSol M f^\insp(t)}{\MSol}\right]^{2/3}
\ee
where $D$ is the distance to the source.

We model the inspiral component from the time the instantaneous frequency
enters the sensitive band of the detector above $f_s$ up to the
commencement of the merger component. Deciding where the boundary between
inspiral and merger lies is somewhat arbitrary. We follow
\cite{fh:waveforms} in making the transition at the point where post-Newtonian
approximations begin to break down.
It is convenient to fix this transition at $t = 0$.
A conservative estimate \cite{fh:waveforms} is that errors in the 2PN
approximation become significant when the instantaneous frequency reaches
\be
f_0 &=& \frac{\MSol}{M} \times 4100\ \Hz
\label{eq:f0}
\ee
so we set the coalescence time $t_c$ of the inspiral in
\cite[eqn. 15.24]{lsc:lsd} by solving $f^\insp(0) = f_0$.

The ringdown component is assumed to be an exponentially damped
sinusoid with constant frequency $f^\ring$ as given in
\cite[eqn. 18.3]{lsc:lsd}. Our amplitude model, adapted
from \cite{kt:300}, is
\be
A^\ring(t) &=& \frac{{\cal A}}{\sqrt{20\pi}}
\frac{\TSol c}{D} \frac{M}{\MSol} e^{-\pi f^\ring (t - t_m)/Q}
\label{eq:Aring}
\ee
where $a$ is the dimensionless spin parameter, $Q = 2(1 -
a)^{-0.45}$ is the quality factor, \be {\cal A} &=& 4 \left[ \frac{\pi
\epsilon} {Q \left[1 - 0.63(1 - a)^{0.3}\right]} \right]^{1/2} \ee and
$\epsilon$ is the fraction of $M$ radiated as gravitational waves
during the ringdown. The factor of $1/\sqrt{20 \pi}$ in
(\ref{eq:Aring}) comes from averaging over orientations and sky
positions. This is essentially the same amplitude model as given in
\cite[eqn. 18.5]{lsc:lsd}.

Our inspiral component has been arranged to terminate at $t = 0$, with
ringdown commencing at $t = t_m$.  Since no analytic models exist for
the merger component, we fit the amplitude and phase functions to
bridge the gap between inspiral and ringdown.  Assuming that the
merger waveform is of the form (\ref{eq:prototype}), a simple way to
connect the inspiral and ringdown waveforms is to require that the
amplitude be continuous to first derivatives, and the phase to be
continuous up to second derivatives (thus ensuring that the
instantaneous frequency is continuous up to first derivatives).  This
gives four conditions that must be satisfied by $f^\merge(t)$ and
$A^\merge(t)$ at $t = 0$ and $t = t_m$, so we model $f^\merge(t)$ and
$A^\merge(t)$ by cubic polynomials.  Since we also require the phase
to be continuous at $t = 0$, we obtain $\phi^\merge(t)$ from the
anti-derivative of $f^\merge(t)$ with an appropriate constant of
integration. We note that phenomenological templates for coalescing
binaries have recently become available \cite{pa:template}, however
our waveforms are qualitatively very similar, and for testing purposes
it is convenient to know the exact form of the instantaneous frequency
and be able to set the precise times of transition from inspiral to
merger to ringdown. Phenomenological templates will be examined 
in future work.

\subsection{Choice of signal parameters}

\begin{figure}[ht]
\includegraphics[width=0.5\textwidth]{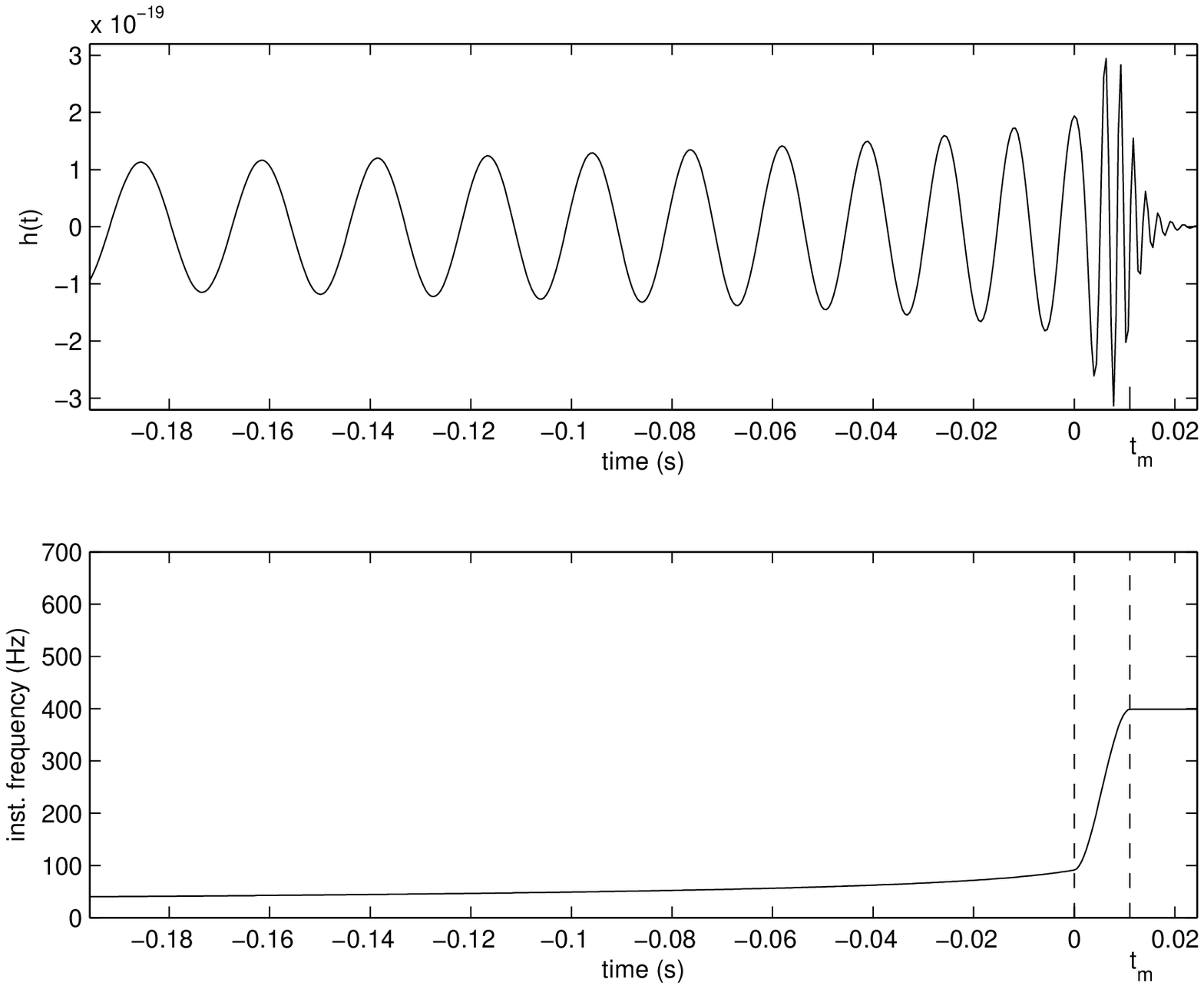}
\includegraphics[width=0.5\textwidth]{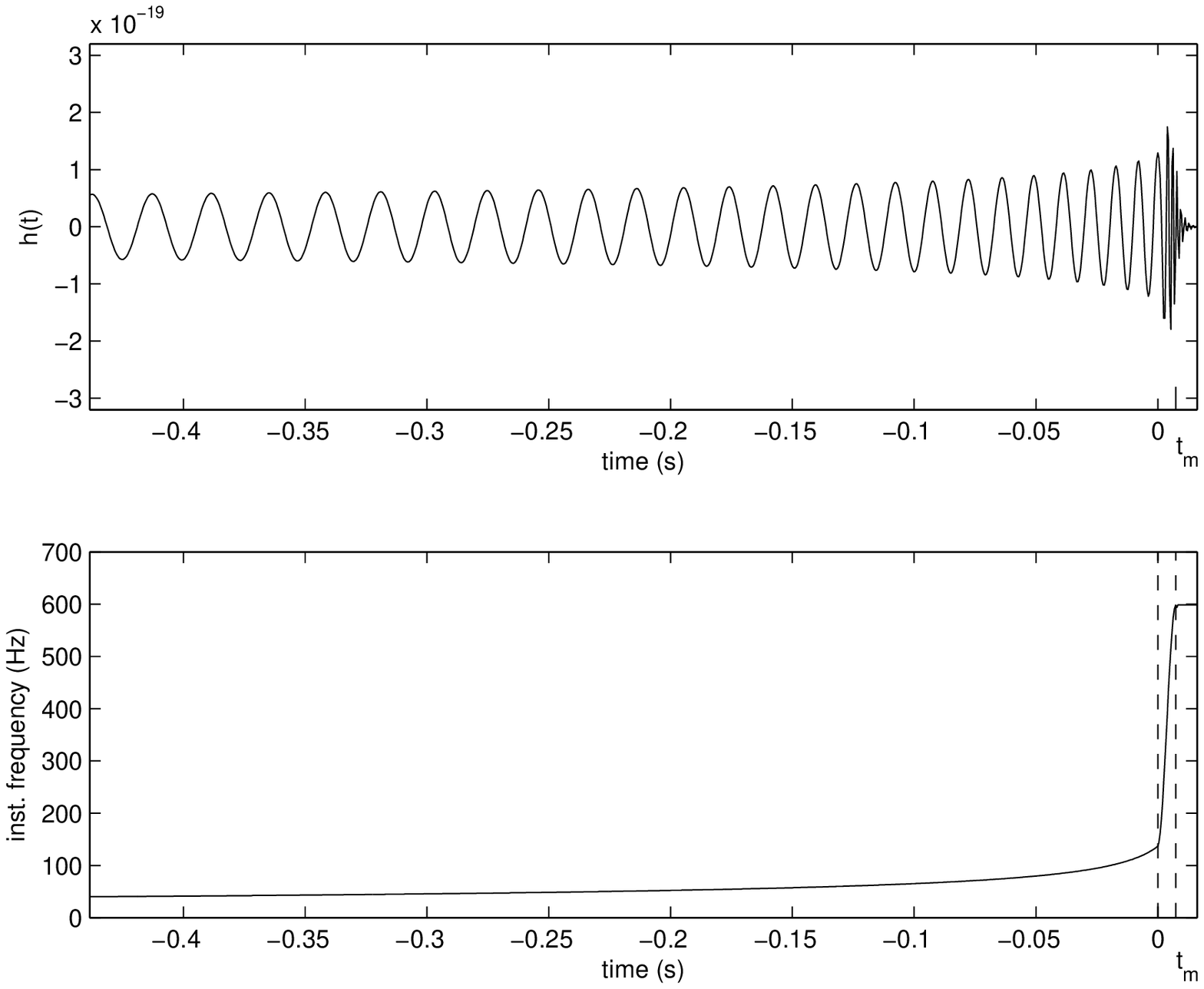}
\phantom{.} \hfill (a) \hfill (b) \hfill \phantom{.}
\caption{\label{fig:gwplot30}$h(t)$ and instantaneous frequency for a
binary coalescence
with masses (a) $m_1 = m_2 = 22.5\ \MSol$ and (b) $m_1 = m_2 = 15\ \MSol$ at a
distance of 1 Mpc.}
\end{figure}

To test detection efficiency we used signals of length $N = 512$, $N =
1024$ and $N = 2048$ sampled at $2048\ \Hz$.  Signals of roughly this
duration are produced by BBH systems with total mass in the
range $20$--$50\ \MSol$.  As most models for the ringdown waveforms
assume equal mass binaries, we will only consider this case.  The
masses used were $m_1 = m_2 = 22.5,\ 15$ and $10$.
Motivated by recent numerical experiments
\cite{fp:evolution,bcckm:binary}, we take $a = 0.7$ and $\epsilon =
0.01$.  While the procedure for producing a merger waveform
is crude, it does produce a signal with frequency and
amplitude characteristics similar to those seen in numerical
relativity simulations.  Figure \ref{fig:gwplot30} shows the strain and
instantaneous frequency for these binary coalescences at a
distance of $1\ \Mpc$ for the $M = 45$ and $30\ \MSol$ cases.


\section{Results}
\label{sec:results}

To use the BP test we first need the distribution of $T^*_\ell$
under $H_0$. There is no analytic expression for the distribution of
the BP statistic so we have used a Monte Carlo simulation to estimate
them.  As our test signals have different lengths we generated three
null distributions, one for each $N$. In each case we generated $10^5$
instances of simulated LIGO noise and calculated $T^*_\ell$ for each of
them with chirplet path lengths $\ell$ drawn from the set $L = \{ 1,
2, 4, 8, 16 \}$.  These random trials give an approximation to the
distributions of $T^*_\ell$ under $H_0$ for each $\ell$.  Using our
empirical distributions we can estimate the $p$-value for an observed
$T^*_\ell$.

\begin{figure}[ht]
(a)
\includegraphics[width=0.5\textwidth]{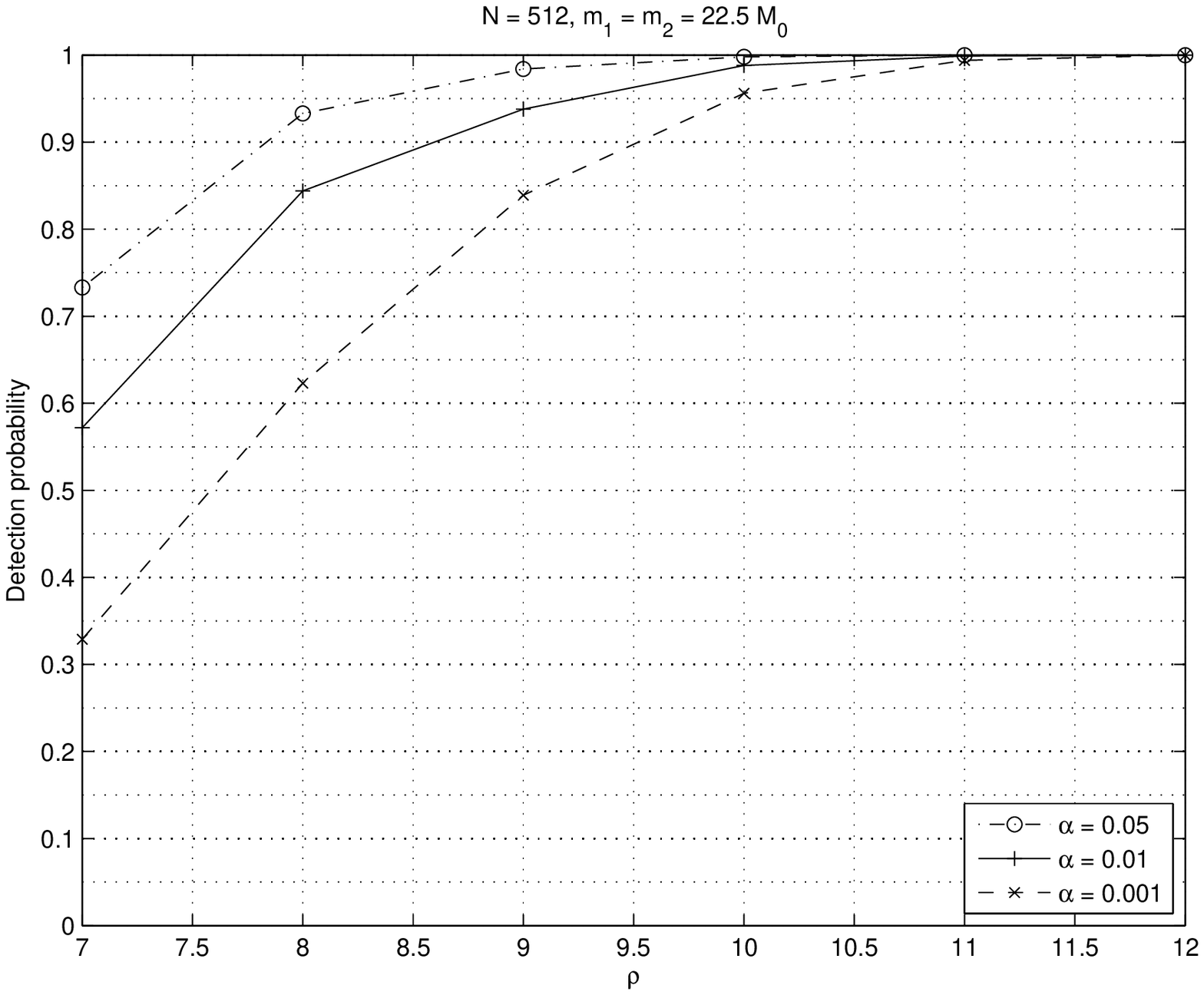}
\includegraphics[width=0.5\textwidth]{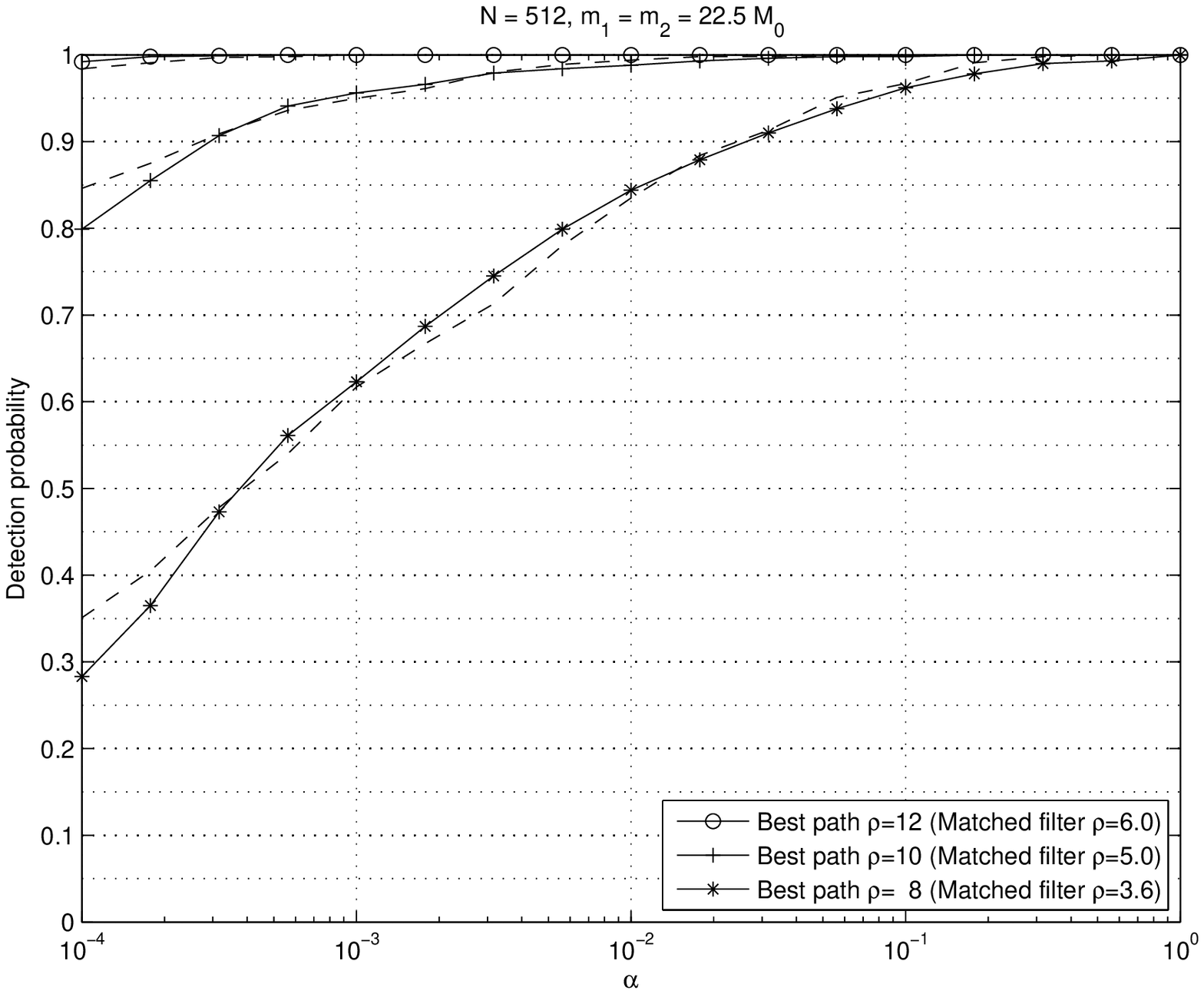}
\\
(b)
\includegraphics[width=0.5\textwidth]{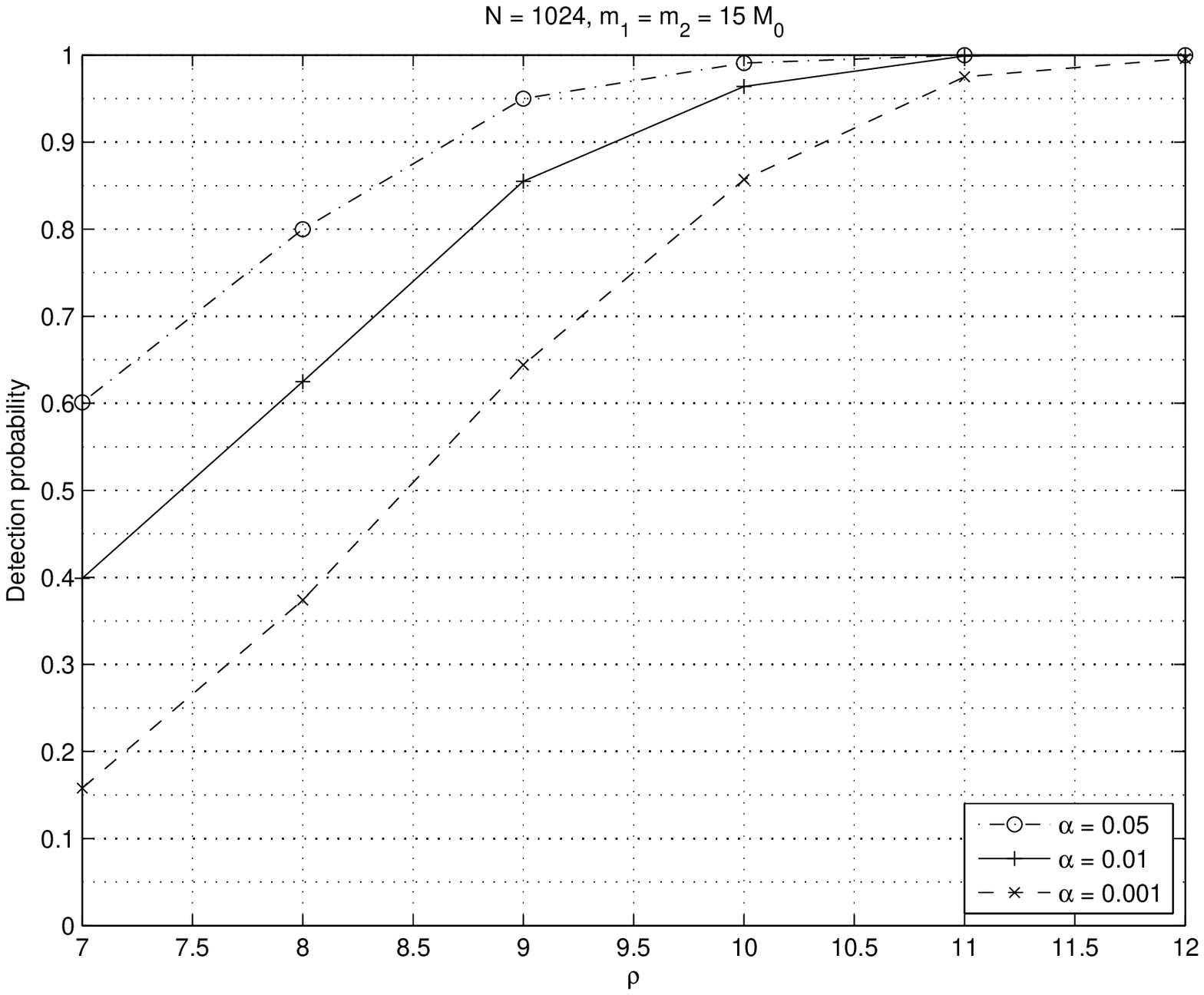}
\includegraphics[width=0.5\textwidth]{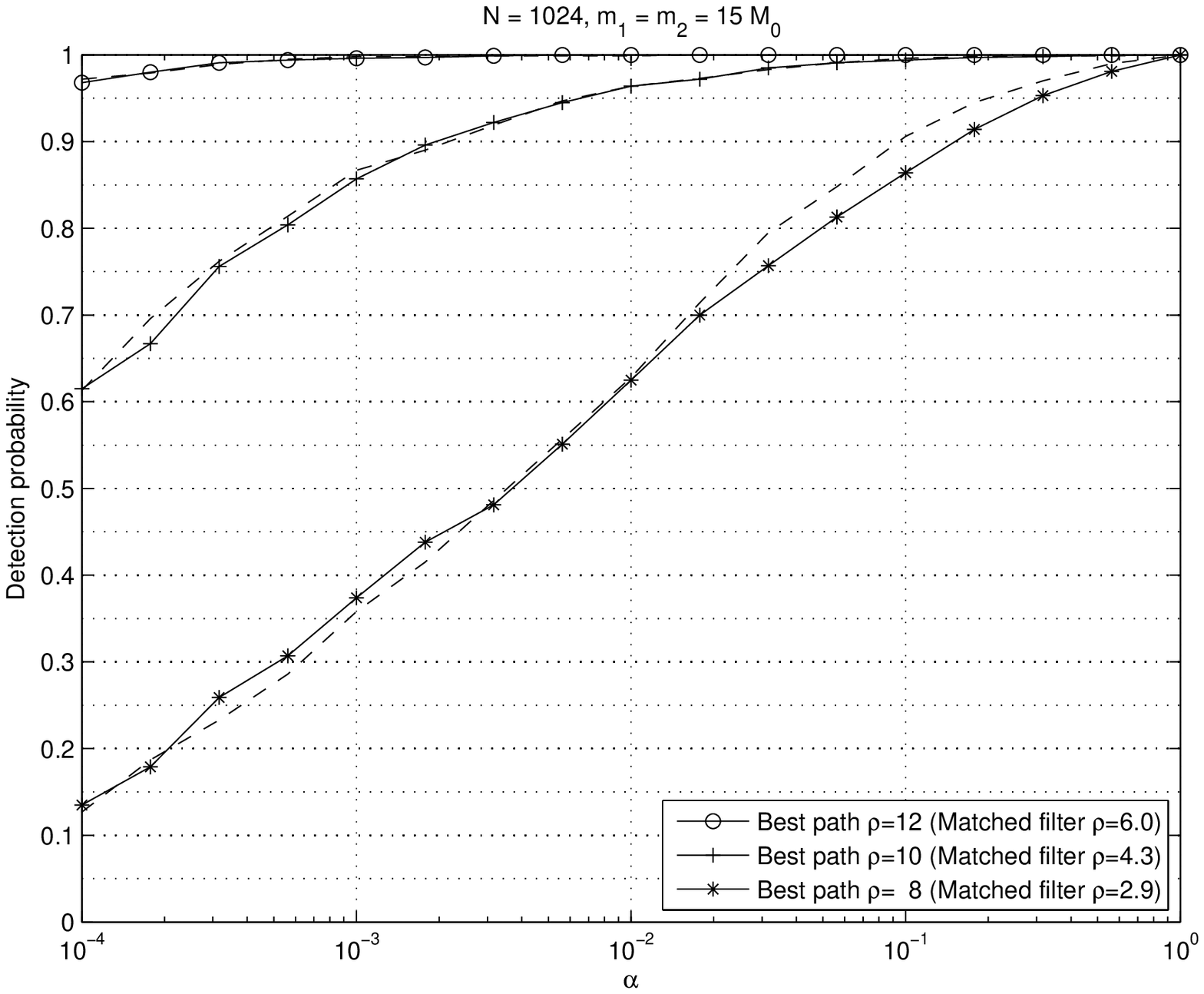}
\\
(c)
\includegraphics[width=0.5\textwidth]{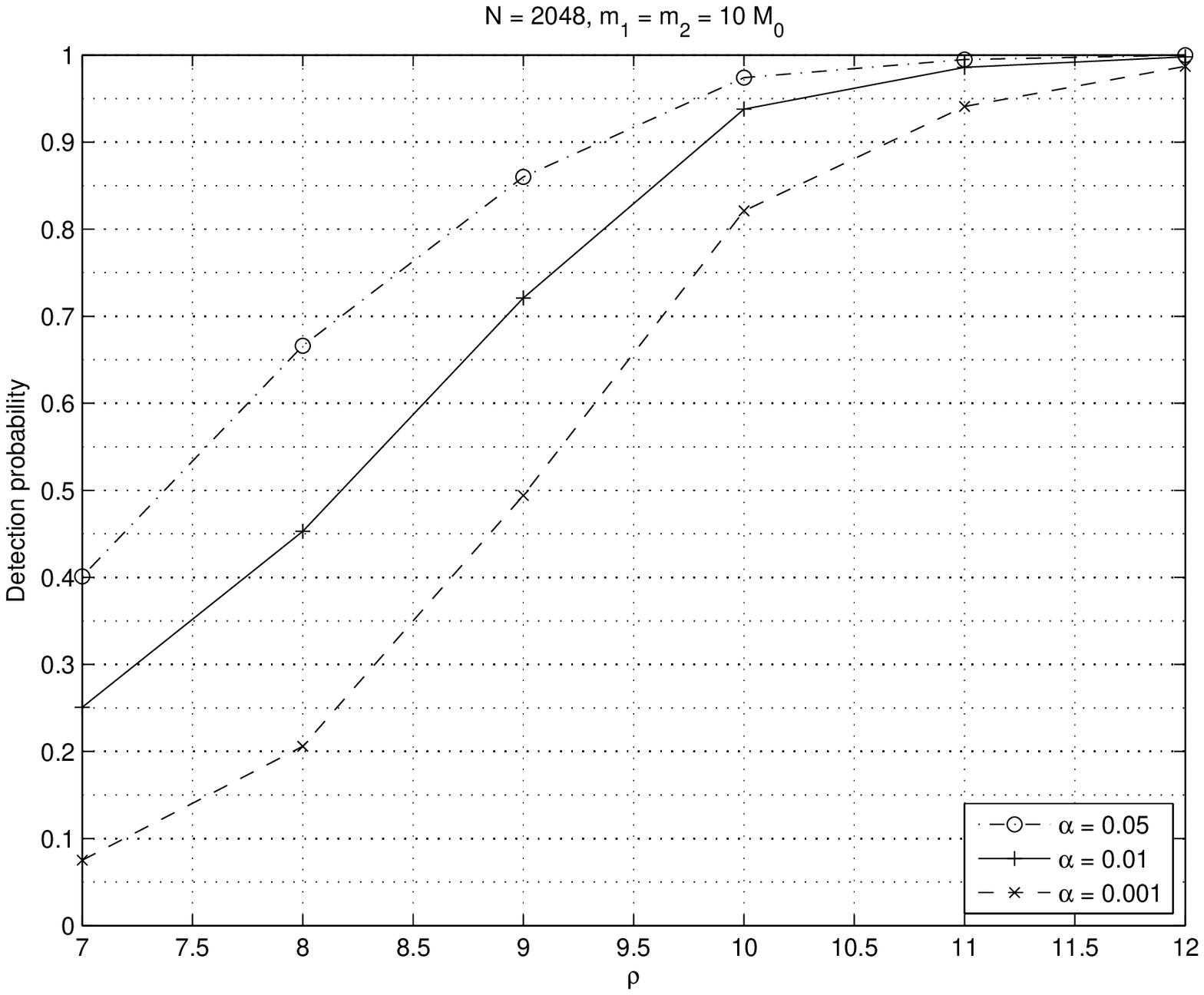}
\includegraphics[width=0.5\textwidth]{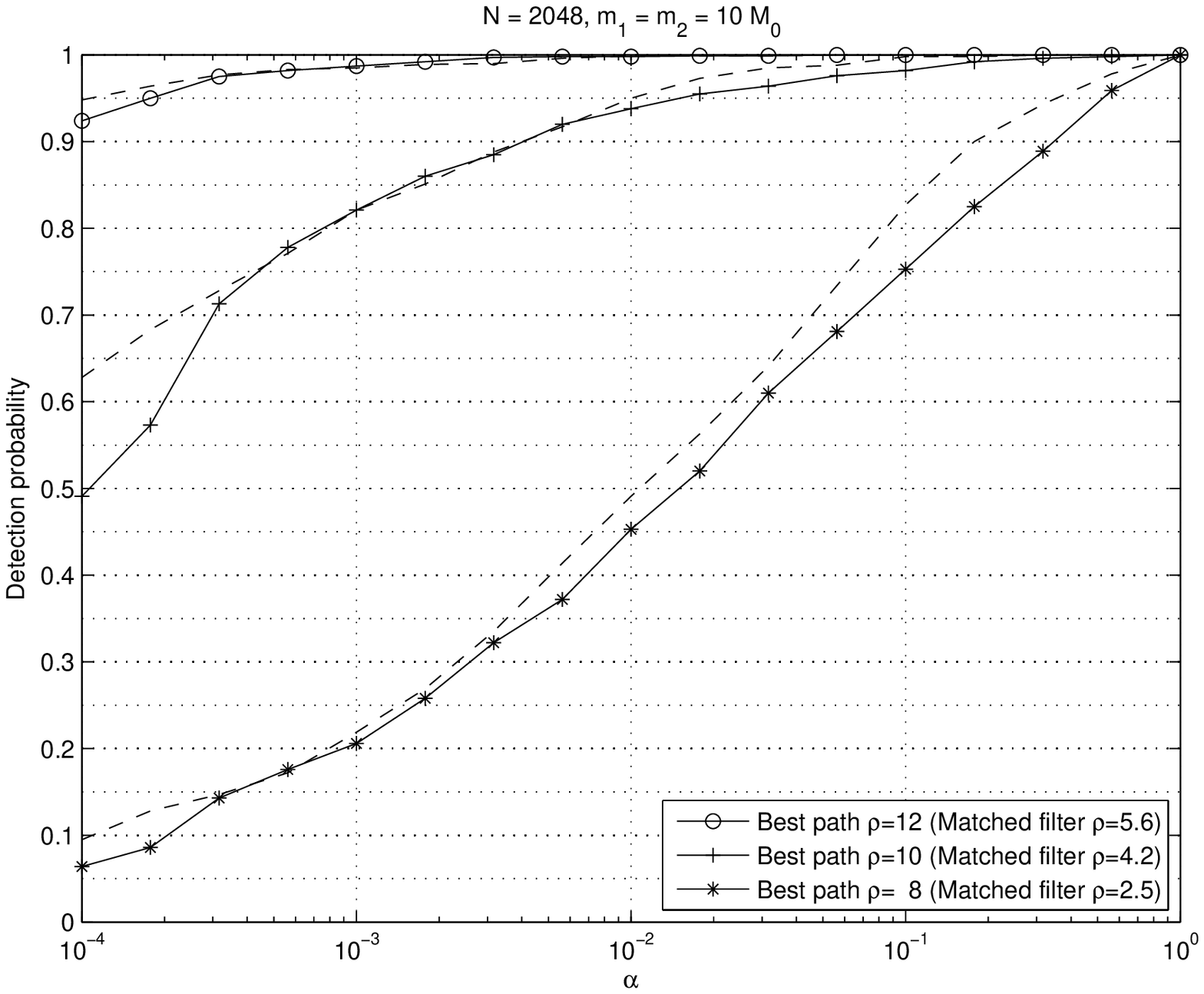}
\caption{\label{fig:gwDPvAlpha45_512}Detection probability as a function of $\rho$
and false alarm probability for BBH coalescences with total mass
(a) $M = 45\ \MSol$ (b) $M = 30\ \MSol$ and (c) $M = 20\ \MSol$.
In each case we give an estimate for the $\rho$ which gives a similar curve using
matched filtering.
}
\end{figure}

To test detection efficiency, we first constructed normalised test signals
using the model described in Section \ref{sec:signal}.  For each
$\rho = 7, 8, 9, 10, 11$ and $12$ we generated $10^5$ instances of
noise and injected the signal at that level. The BP statistic $T^*_\ell$
was calculated, as was $p^*$, and we determined the detection
probability for a given $\alpha$ by counting
the number of $p^* \le \alpha$.  Figure \ref{fig:gwDPvAlpha45_512}
gives the detection probabilities as a function of $\alpha$
(the Receiver Operating Characteristic curve)
for $\rho = 8$, $10$ and $12$.  For comparison, we also give an
ROC curve obtained using matched filtering to detect the signal.  For
these curves $\rho$ has been chosen to give a
good match to the ROC curve obtained via the BP test. From this it can be seen that
the BP statistic is about half as sensitive as matched
filtering. Since the distance $D$ to the source is inversely
proportional to the overall signal amplitude, we can consider the BP
test to have a seeing distance about half that of matched filtering.

In Figure \ref{fig:gwDPvAlpha45_512} we also give the detection
probability as a function of $\rho$ (or equivalently, inverse distance to
the source) for $\alpha$ = $0.05$, $0.01$ and $0.001$.
The corresponding distances at $\rho = 10$ are $D = 100\
\Mpc$ for $M = 45\ \MSol$, $D = 80\ \Mpc$ for $M = 30\ \MSol$ and $D =
65\ \Mpc$ for $M = 20\ \MSol$.  This shows that, for example, at a
false alarm probability of $\alpha = 0.001$ we can see an event out to
$\sim 100\ \Mpc$ with a false dismissal probability of about $10\%$.
Note that since we have averaged the signal amplitude over sky
positions and orientations, an
optimally aligned and positioned source could be detected much farther
away.

In the above comparison we are injecting a known signal into noise
and using the same (normalised) signal as our template for matched
filtering. Real signals in interferometer data will have unknown
parameters, and a bank of templates using discrete values of the
parameters (mass, spin etc) is needed to cover the range of physically
plausible coalescences. Since a real signal has parameters drawn from
a continuum there will usually be some degree of mismatch between the
signal and templates in the bank. As such, the comparison above is very conservative in comparing the BP test
with the most favourable matched filter detection scenario, one which
is unlikely to be attained in practise.
A more realistic benchmark is obtained by examining the performance of
the BP test when the signal parameters are chosen at random from a
range of values.  Here we present a comparison of the BP test with detection
via a bank of templates, and with another method employed in searches
for unmodelled signals, the {\em excess power} statistic \cite{abcf:ep}.

We first created a bank of templates using discrete values for
the parameters. Although our complete signal model contains a large
number of free parameters, for simplicity we chose to only vary $m_1$,
$m_2$ and $a$, as these have the greatest effect on the waveform.  For
the same reason we have used equal spacing in all parameters, rather
than attempting to construct a template bank spaced to give equal
overlap between adjacent templates. While methods exist to construct
optimally-spaced template banks, our templates have the additional
complication of including merger and ringdown components.

For each of the signal lengths $N = 512$, $1024$ and $2048$ we
generated a bank of normalised templates using the criteria that
\begin{enumerate}
\renewcommand{\labelenumi}{\arabic{enumi}.}
\item The range of masses $m_1$ and $m_2$ is chosen
so that the length of the signals range from $N/2$ to $N$ samples.
\item The spin ranges from $a = 0.18$ to $0.98$.
\item The spacing between masses and spins is chosen so that the minimal match
of a signal with parameters drawn from the range of parameters is
at least $0.97$.
\end{enumerate}

\begin{figure}[ht]
\includegraphics[width=0.5\textwidth]{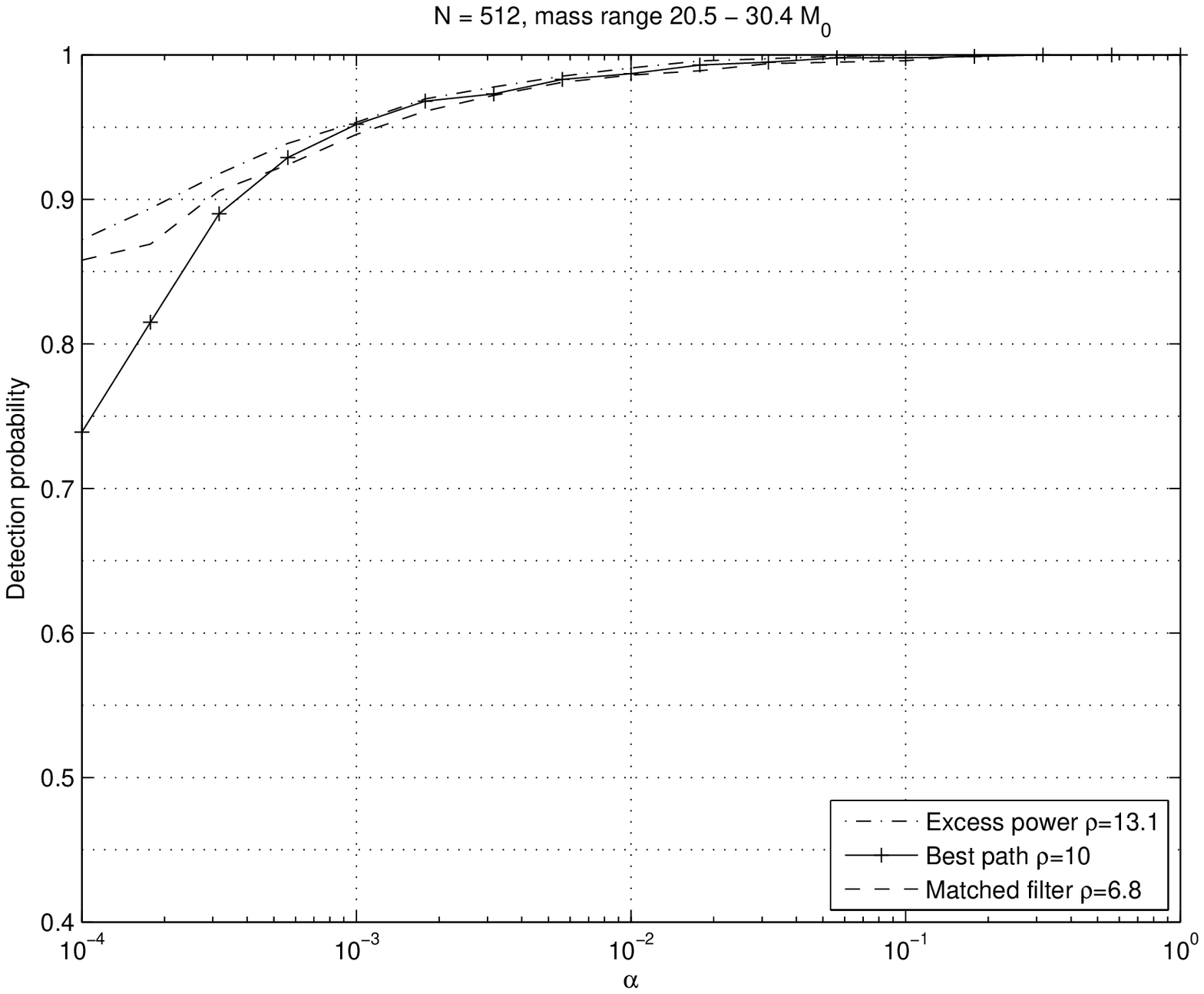}
\includegraphics[width=0.5\textwidth]{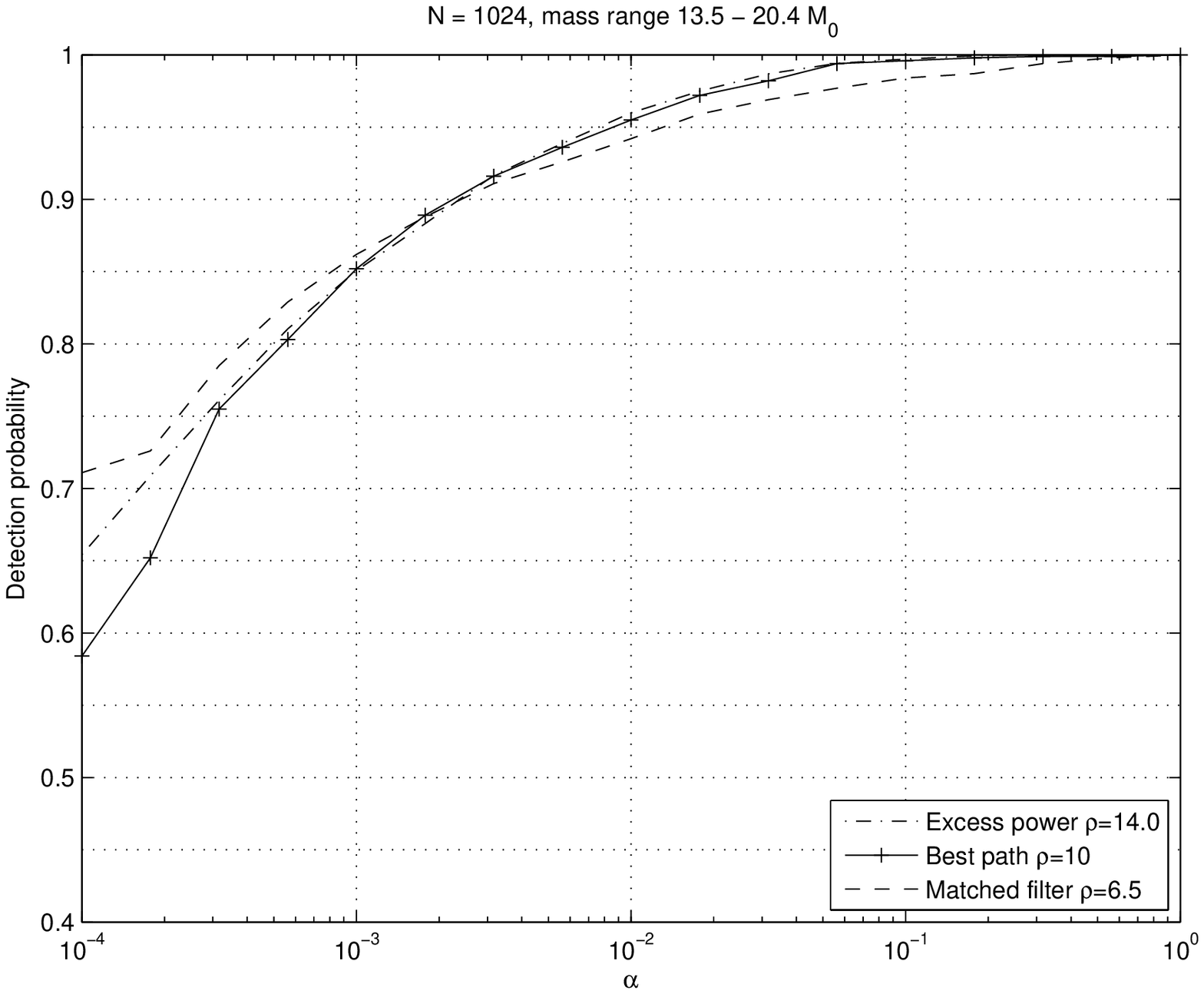}
\\
\phantom{.} \hspace{2.8cm} (a) \hspace{6cm} (b)\\
\begin{center}
\includegraphics[width=0.5\textwidth]{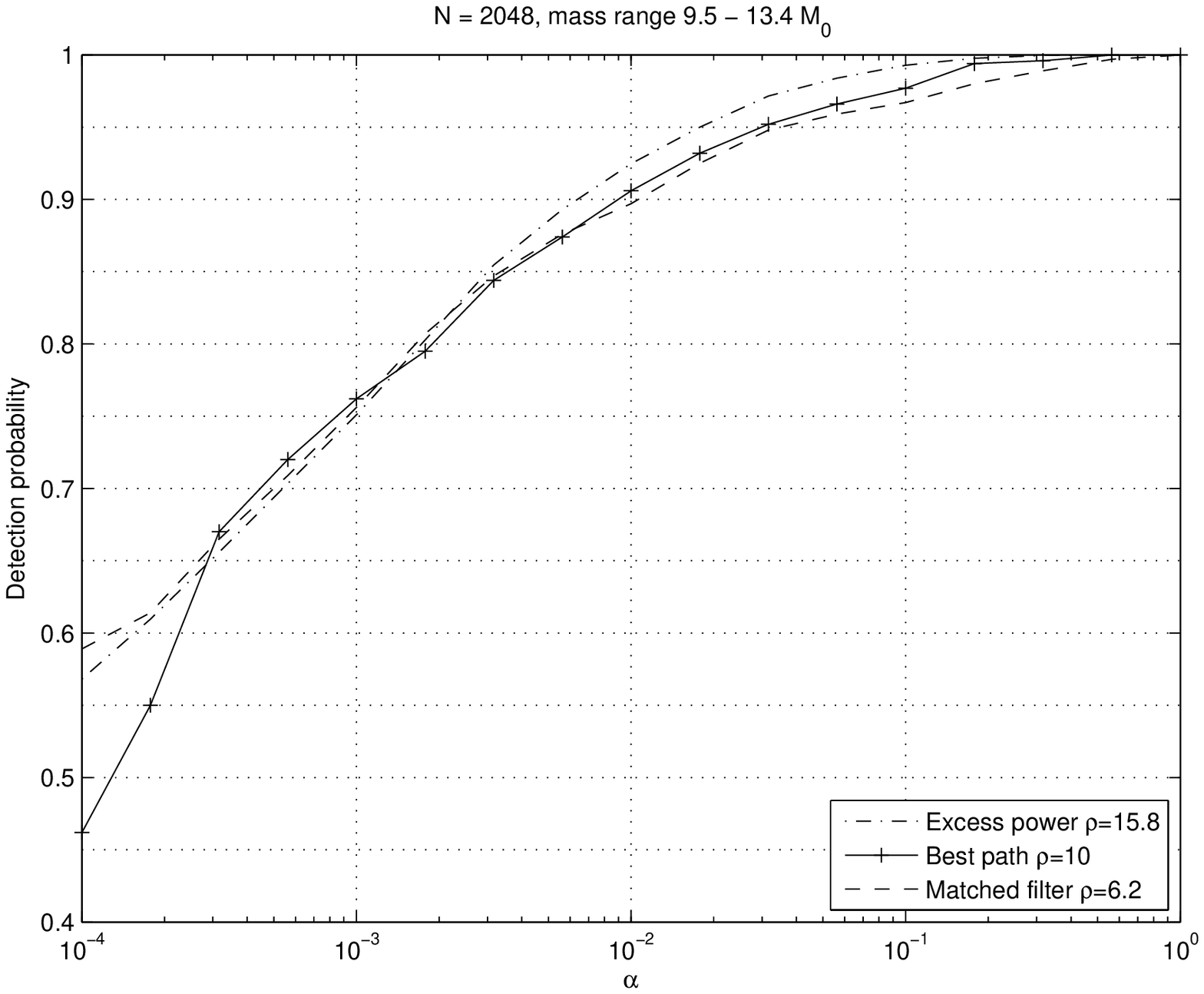}
\\
(c)
\end{center}
\caption{\label{fig:gwCSPvBANK45_512}Detection probability as a function of
false alarm probability for random signals in the mass ranges (a) $20.5$--$30.4\ \MSol$
(b) $13.5$--$20.4\ \MSol$ and (c) $9.5$--$13.4\ \MSol$.
In each case we give an estimate for the $\rho$ which gives a similar curve using
matched filtering with a bank of templates and the excess power statistic.}
\end{figure}

For each $N$ we then generated $1000$ test signals with mass and spin
parameters drawn at random from the appropriate range, and injected
them into simulated LIGO noise with $\rho = 10$. The resulting data
was used to calculate a BP statistic for each
segment. Comparing the BP statistics with the empirical null
distribution as above, we obtained the ROC curves
shown in Figure \ref{fig:gwCSPvBANK45_512}.  Searching for the same
signals via matched filtering, we found that the ROC curves matched
well when the signals were injected with $\rho$ around $6.5$ -- in other
words, the BP test sees about $2/3$ as far as the template bank.

We performed a similar analysis using the excess power statistic,
which is optimal when the only known features of the signal are the
duration and bandwidth \cite{abcf:ep}. The excess power statistic is
simply the power $\inner{u}{u}$ calculated using (\ref{eq:inner})
where the integration is performed over the bandwitch of the expected
signals, taken to be $40$--$1024$ Hz in this instance. Under $H_0$
this has a $\chi^2$ distribution with degrees of freedom
twice the number of frequency bins. Using the excess power statistic,
we found that the ROC curves matched those of the BP test in Figure
\ref{fig:gwCSPvBANK45_512} well when $\rho$ was around $14$.


\section{Conclusion}

Chirplet path pursuit has previously been shown to be effective at
detecting a broad class of chirp-like but otherwise unmodelled signals
in coloured noise \cite{chc:theory}.  In this paper we have
demonstrated that the method can be successfully applied to the
problem of detecting test signals with similar characteristics to
those expected from binary black hole coalescence.  The method is able
to detect a range of signals of modest strength hidden in
simulated LIGO noise, and exhibits somewhat better statistical power
than the excess power test.

As with other methods for detecting bursts, in real LIGO noise there
is the difficulty of distinguishing genuine gravitational wave signals
from instrumental and environmental events. For matched filter
searches the $\chi^2$ discriminator can be used to reject signals that
do not have the correct distribution of power across frequency bands,
however this requires that the gravitational waveform be known
\cite{ba:chi2}. This discriminator is not applicable to chirplet path
pursuit since the signal is not known and we do not impose any assumptions on the distribution of
power. Instead we would rely on the methods being
employed in current searches: requiring events to be coincident across
multiple detectors, vetoing events based on environmental channels,
and testing if waveforms measured in different detectors are
consistent \cite{lsc:bursts,lsc:burstsS4}.

As expected for a non-parametric method, chirplet path pursuit is not
as sensitive as matched filtering using a template bank, nevertheless
our comparison shows that the method has similar effectiveness to
matched filtering for a signal that is roughly $1.5$ times as
strong. Significantly, since the method is sensitive to a wide range of
chirp-like signals, an exact model of the signals to be detected is
not necessary. This makes the method particularly of interest in
situations where the signal is unmodelled or poorly modelled, as is
the case for the late inspiral and merger components of intermediate
mass black hole coalescences.


\ack
\comment{
The authors gratefully acknowledge the support of the United States
National Science Foundation for the construction and operation of the
LIGO Laboratory and the Science and Technology Facilities Council of the
United Kingdom, the Max-Planck-Society, and the State of
Niedersachsen/Germany for support of the construction and operation of
the GEO600 detector. The authors also gratefully acknowledge the support
of the research by these agencies and by the Australian Research Council,
the Council of Scientific and Industrial Research of India, the Istituto
Nazionale di Fisica Nucleare of Italy, the Spanish Ministerio de
Educacion y Ciencia, the Conselleria d'Economia Hisenda i Innovacio of
the Govern de les Illes Balears, the Scottish Funding Council, the
Scottish Universities Physics Alliance, The National Aeronautics and
Space Administration, the Carnegie Trust, the Leverhulme Trust, the David
and Lucile Packard Foundation, the Research Corporation, and the Alfred
P. Sloan Foundation.
}
EC was partially supported by National Science Foundation grants DMS
01-40698 (FRG) and ITR ACI-0204932. We thank Warren Anderson
for supplying his Maple code for simulating BBH coalescence.
This article has LIGO Document Number LIGO-P080017-00-Z.


\section*{References}


\begin{thebibliography}{10}
\expandafter\ifx\csname url\endcsname\relax
  \def\url#1{{\tt #1}}\fi
\expandafter\ifx\csname urlprefix\endcsname\relax\def\urlprefix{URL }\fi
\providecommand{\eprint}[2][]{\url{#2}}

\bibitem{ligo:web}
\verb+http://www.ligo.caltech.edu+

\bibitem{dis:comp}
Damour T, Iyer B~R and Sathyaprakash B~S 2001 {\em Phys. Rev. D\/} {\bf 60}
  044023

\bibitem{acst:spin}
Apostolatos T~A, Cutler C, Sussman G~J and Thorne K~S 1994 {\em Phys. Rev. D\/}
  {\bf 49} 6274

\bibitem{ss:templates}
Sathyaprakash B~S and Schutz B~F 2003 {\em Class. Quantum Grav.\/} {\bf 20}
  S209--S218

\bibitem{gbcclpv:event}
Gair J~R, Barack L, Creighton T, Cutler C, Larson S~L, Phinney E~S and
  Vallisneri M 2004 {\em Class. Quantum Grav.\/} {\bf 21} S1595--S1606

\bibitem{bcv:detecting}
Buonanno A, Chen Y and Vallisneri M 2003 {\em Phys. Rev. D\/} {\bf 67} 024016

\bibitem{pbcv:physical}
Pan Y, Buonanno A, Chen Y and Vallisneri M 2004 {\em Phys. Rev. D\/} {\bf 69}
  104017

\bibitem{bcptv:detecting}
Buonanno A, Chen Y, Pan Y, Tagoshi H,  and Vallisneri M 2005 {\em Phys. Rev.
  D\/} {\bf 72} 084027

\bibitem{lg:search}
Goggin L 2006 {\em Class. Quantum Grav.\/} {\bf 23} S709--S713

\bibitem{fp:evolution}
Pretorius F 2005 {\em Phys. Rev. Lett.\/} {\bf 95} 121101

\bibitem{bcckm:binary}
Baker J~G, Centrella J, Choi D~I, Koppitz M and {van Meter} J 2006 {\em Phys.
  Rev. D\/} {\bf 73} 104002

\bibitem{ab:tf}
Anderson W~G and Balasubramanian R 1999 {\em Phys. Rev. D\/} {\bf 60} 102001

\bibitem{abcf:ep}
Anderson W~G, Brady P~R, Creighton J~D~E and Flanagan {\'{E}}~{\'{E}} 2001 {\em
  Phys. Rev. D\/} {\bf 63} 142003

\bibitem{js:time}
Sylvestre J 2002 {\em Phys. Rev. D\/} {\bf 66} 102004

\bibitem{km:wb}
Klimenko S and Mitselmakher G 2004 {\em Class. Quantum Grav.\/} {\bf 21}
  S1819--S1830

\bibitem{sc:bursts}
Chatterji S 2005 {\em The search for gravitational wave bursts in data from the
  second {LIGO} science run\/} Ph.D. thesis Massachusetts Institute of
  Technology

\bibitem{cp:best}
{C}hassande{-}{M}ottin {\'{E}} and Pai A 2006 {\em Phys. Rev. D\/} {\bf 73}
  042003

\bibitem{lsc:burstsS4}
{A Abbott {\em et al}} 2007 {\em Class. Quantum Grav.\/} {\bf 24} 5343--5369

\bibitem{chc:theory}
Cand\`{e}s E~J, Charlton P~R and Helgason H 2008 {\em Appl. Comput. Harmon.
  Anal.\/} {\bf 24} 14--40

\bibitem{jp:fct}
Jenet F~A and Prince T~A 2000 {\em Phys. Rev. D\/} {\bf 62} 122001

\bibitem{hj:shortest}
Joksch H~C 1966 {\em J. Math. Anal. Appl.\/} {\bf 14} 191--197

\bibitem{bh:false}
Benjamini Y and Hochberg Y 1995 {\em J. R. Statist. Soc. B\/} {\bf 57} 289--300

\bibitem{glpps:gwa}
Grishchuk L~P, Lipunov V~M, Postnov K~A, Prokhorov M~E and Sathyaprakash B~S
  2001 {\em Physics-Uspekhi\/} {\bf 44} 1--51

\bibitem{fh:waveforms}
Flanagan {\'{E}}~{\'{E}} and Hughes S 1998 {\em Phys. Rev. D\/} {\bf 57} 4535

\bibitem{lsc:lsd}
{The LIGO Scientific Collaboration} {LAL} {S}oftware {D}ocumentation
  \url{http://www.lsc-group.phys.uwm.edu/lal/slug/nightly/doc/lsd-nightly.pdf}

\bibitem{kt:300}
Thorne K~S 1987 {\em 300 Years of Gravitation\/} ed Hawking S~W and Israel W
  (Cambridge University Press)

\bibitem{pa:template}
{P Ajith {\em et al}} 2007 {\em Class. Quantum Grav.\/} {\bf 24} S689--S699

\bibitem{ba:chi2}
Allen B 2005 {\em Phys. Rev. D\/} {\bf 71} 062001

\bibitem{lsc:bursts}
{A Abbott {\em et al}} 2005 {\em Phys. Rev. D\/} {\bf 72} 122004

\bibitem{ec:chirplets}
Cand\`{e}s E~J 2002 Multiscale chirplets and near-optimal recovery of chirps
  Tech. rep. Stanford University

\end{thebibliography}
\providecommand{\newblock}{}

\appendix
\section*{Appendix}
\setcounter{section}{1}

In this appendix we present the scheme used for calculating chirplet
coefficients of discretised data.
The data $u[n] = u(n\Delta t)$ is discretely sampled at $N = 2^S$
intervals of duration $\Delta t$. Notionally, this discretises the time-frequency
plane into points $(t_i, f_k)$ where $t_i = i\Delta t$ and $f_k =
k\Delta f = k/(N\Delta t)$. Points in the time-frequency plane are
considered to be vertices in a directed graph where the weight of the
arc connecting two vertices is given by the local correlation of
$u(t)$ with the corresponding chirplet.

Consider chirplets supported on the interval $[0, 2^{-s}T)$.  At
scale $s$, $0 \le s < S$ this interval has length $N_s = 2^{-s}N$ samples. While there
are many ways to discretise chirplets on this interval, it is
convenient to choose the spacing of the frequency parameter to correspond with
the bins of a discrete Fourier transform, and choose the spacing of the chirp
parameter so that at the end of the interval the instantaneous
frequency has changed by a whole number of bins. Thus our dictionary
of chirplets is indexed by scale index $s$, frequency
index $k$ and chirp index $l$, and the (unnormalised) discrete chirplet is given by
\begin{eqnarray}
c_{s,k,l}[n] &=& e^{{i2\pi \phi_{s,k,l}[n]}} \qquad 0 \le n < N_s,\ \ 0 \le k \le N/2
\end{eqnarray}
where the phase is
\begin{eqnarray}
\phi_{s,k,l}[n] &=& k\frac{n}{N} + l\frac{n^2}{2 N N_s}.
\end{eqnarray}
The discretised instantaneous frequency is
\begin{eqnarray}
\dot\phi_{s,k,l}[n] &=& k + l\frac{n}{N_s}.
\end{eqnarray}
Such a chirplet has initial frequency $k\Delta f\ \Hz$ and rises to frequency $(k+l)\Delta f\ \Hz$
at a rate of $l \Delta f/(N_s\Delta t)\ \mbox{Hz s}^{-1}$.
Since we only deal with real signals, the range of the chirp index $l$ is chosen to
restrict the chirplets to non-negative frequencies up to Nyquist, thus $-k \le l \le N/2 - k$.

In general, the inner product (\ref{eq:inner}) for a noise process with covariance matrix $\Sigma$
is $u^*\Sigma^{-1}v$, where $u^*$ is the conjugate transpose of $u$. For our
noise model the Fourier matrices $F_{mn} = e^{-i2\pi m n/N}$ diagonalise $\Sigma$, and
so $\inner{u}{v} = u^*F^*D^{-1}Fv = \ft{u}^*D^{-1}\ft{v}$ where
$D = \mbox{diag}(\sigma^2_0, \sigma^2_1, \ldots, \sigma^2_{N-1})$ and $\sigma^2_k = \langle|\ft{n}_k|^2\rangle$
are the eigenvalues of $\Sigma$.
Calculating $\inner{u}{c}$ for a chirplet supported on a dyadic interval $I_{s,j} =
[j 2^{-s}T, (j+1) 2^{-s}T)$ is equivalent to calculating the inner product
of $u(t+j 2^{-s}T)$ with a chirplet supported on $[0, 2^{-s}T)$.
As the time index of the first sample in $I_{s,j}$ is $j N_s$, let
$u_{s,j} = (u[jN_s], u[jN_s+1],\ldots, u[(j+1)N_s-1])$
be the samples of $u(t)$ restricted to $I_{s,j}$.
Then to find $\inner{u}{c}$ we pad $u_{s,j}$ and $c$ to length $N$ with zeroes and FFT.
In discrete form, the inner product then reduces to
\begin{eqnarray}
\inner{u}{c} &=& \frac{\Delta t}{N}\sum_{n\,=\,0}^{N-1} \frac{\ft{u}^*_{s,j}[n] \tilde{c}[n]}{S[n]}
\end{eqnarray}
where $S[n] = S(n\Delta f)$. If $c$ has indices $k$, $l$ then after
normalising, $|\inner{u}{c}|^2/|\inner{c}{c}|^2$ is the
weight of the arc connecting $(t_{jN_s}, f_k)$ to $(t_{(j+1)N_s},
f_{k+l})$.

To calculate the BP statistic we must find the total weight
of connected, non-overlapping chirplet paths in the time-frequency
plane starting at $t = 0$ and ending at $t = T$. To keep the number of arcs manageable we further restrict
our chirplet paths to those supported on a {\em recursive dyadic
partition} (RDP) of $I$ constructed using the following definition
\cite{ec:chirplets}:
\begin{enumerate}
\item[1.] The trivial partition ${\cal P} = \{ I \}$ is an RDP.
\item[2.] If ${\cal P} = \{ I_1, I_2, \ldots, I_p \}$ is an RDP, then so is the
partition obtained by splitting any interval $I_j$ into two adjacent dyadic intervals.
\end{enumerate}
This means that, for example, that $\{ [0, 1/4), [1/4, 1/2), [1/2, 1)
\}$ is a recursive dyadic partition of $[0, 1)$, but $\{ [0, 1/4),
[1/4, 1) \}$ is not. The total weight of a chirplet path $P = \{ c_1,
c_2, \ldots, c_p \}$ supported on ${\cal P} = \{ I_1, I_2,
\ldots, I_p \}$ is then
\begin{eqnarray}
T_P &=& \sum_{p} \frac{|\inner{u}{c_p}|^2}{|\inner{c_p}{c_p}|^2}
\end{eqnarray}

\end{document}